\documentstyle[12pt]{article}
\newcommand{\ha}{\mbox{\small$\frac{1}{2}$}}
\newcommand{\qu}{\mbox{\small$\frac{1}{4}$}}
\newcommand{\na}{(-)^{\bar a}}

\newcommand{\s}[1]{{\sf #1}}
\newcommand{\B}[1]{{\bf #1}}

\newcommand{\IR}{{\rm I}\hspace{-0.18em}{\rm R}}
\newcommand{\IM}{{\rm I}\hspace{-0.18em} {\rm I}\hspace{-0.31em}{\rm M}}

\newcommand{\sumab}{\sum \sum_{\hspace*{-1.2em} a\ <\ b}}

\newcommand{\inta}{\hspace*{-.2em}\int \hspace*{-.3em}}
\newcommand{\intab}{\hspace*{-.2em}\int \hspace*{-.4em}\int \hspace{-.3em}}

\newcommand{\lr}[1]{\sqrt{\vphantom{\dot z^2_a}\dot z^2_{#1}}}

\title{FOKKER--TYPE CONFINEMENT MODELS FROM EFFECTIVE LAGRANGIAN
IN CLASSICAL YANG--MILLS THEORY}

\author{A.~DUVIRYAK\thanks{E--mail: duviryak@omega.icmp.lviv.ua}
\vspace{1ex}\\ 
{\em Institute for Condensed Matter Physics}\\
{\em of Ukrainian National Academy of Sciences},\\
{\em 1 Svientsitskyj Street, UA-290011, Lviv, Ukraine}}

\date{}

\begin{document}
\maketitle
\begin{abstract}
Abelian potentials of pointlike moving sources are obtained from the 
nonstandard theory of Yang--Mills field. They are used for the construction 
of the time-symmetric and time-asymmetric Fokker-type action integrals 
describing the dynamics of two-particle system with confinement interaction. 
The time-asymmetric model is reformulated in the framework of the 
Hamiltonian formalism. The corresponding two-body problem is reduced to 
quadratures. The behaviour of Regge trajectories is estimated within the
semiclassical consideration.
\vspace{1ex}
\end{abstract}
\large
\section{Introduction}

	Potential models of hadrons originate from the quantum
chromodynamics (QCD), but they are not rigorously deducible from
the theory. Rather, these models are substantiated by various approximate
approaches and estimates in QCD.$^{\ref{Sim94}}$ Various models have their
own areas of application. In particular, the linear potential which follows
from area law in the lattice approximation of QCD, describes, by
construction, the static interaction of quarks.$^{\ref{S-F88}}$ Thus it can
lawfully be exploited only in nonrelativistic potential 
models.$^{\ref{L-S89}}$

	The description of light meson spectroscopy needs the
development of appropriate relativistic models. They frequently are
built as single-particle wave equations$^{\ref{FKR71}-\ref{HLS94}}$
which is not satisfactory. Actually, mesons should be treated as composite
two--quark relativistic systems. The reliable basis
for this purpose is the relativistic direct interaction theory (RDIT)
presented by various approaches and formalisms,$^{\ref{Bar82}}$ such as
Lagrangian formalism with higher derivatives,$^{\ref{GKT87}}$
relativistic Hamiltonian mechanics,$^{\ref{L-S78},\ref{Pol89}}$
canonical formalism with constraints,$^{\ref{Fir87}}$
Fokker-type action formalism$^{\ref{H-N74}}$ etc.

	Given a nonrelativistic potential, RDIT determines the general
structure of its relativistic counterpart. In so going the great
arbitrariness arises in the choice of concrete relativistic interaction.
Consequently, the variety of relativistic potential models has appeared
in the literature.$^{\ref{Tak79},\ref{Roh79}}$
Each of them has its own advantages and areas of 
application, but these models are not substantiated by QCD better than 
nonrelativistic models.

	A possible way to substantiate relativistic direct interactions
leads through classical field theory. Especially, we mean the Fokker-type
action formalism which, among other approaches to RDIT, is most closely
related to this theory. There exists the class of Fokker actions which
correspond to particle interactions via linear fields, such as scalar,
vector, and other tensor fields.$^{\ref{Hav71}}$ These actions
are built on the solutions to relevant wave equations. In the nonrelativistic
limit they lead to the same Coulomb (or Yukawa) potential.

	Working within this scope for the confinement case, one could try
to proceed from some solution to classical Yang--Mills equations. But no
such solutions leading to confining potentials are known in the literature.
Moreover, they are believed to not exist due to the essentially quantum
nature of confinement. This is concerned with standard Yang--Mills theory
while there exist various nonstandard theories which involve effective
Yang--Mills fields arising from QCD. These theories may be used as sources
of confining potentials.

	In the present paper we find the relation between certain nonstandard
classical theory of Yang--Mills field and the Fokker-type confinement model.
The former is developed in Ref. \ref{A-A84}. This theory describes non-Abelian
gauge field averaged over quantum fluctuations. It is based on the
effective Lagrangian obtained from the study of infrared behaviour of
gluon Green's functions in QCD.$^{\ref{AAB82}}$ Field equations following
from this Lagrangian are of 4th order, and some static non-Abelian
solutions to them have been used in a sort of bag confinement 
model.$^{\ref{Ale88}}$

	Here we obtain from this theory the Abelian retarded and time-%
symmetric potentials of moving pointlike source. Both of them are of
confining type and reduce in the nonrelativistic limit to the linear
potential. Then, using these potentials, we construct the time-asymmetric
and time-symmetric Fokker-type actions. The latter is already known in
the literature. Two equivalent versions of this action have been
proposed by Rivacoba$^{\ref{Riv84}}$ and Weiss$^{\ref{Wei86}}$. It is
noteworthy that both the authors proceeded from general preliminaries
of RDIT, without referring to field--theoretical interpretation of particle
interaction.

	The time--symmetric action leads to difference--differential 
equations of motion which are difficult to deal with. The only circular--%
orbit solutions to these equations are found in Refs. \ref{Riv84} and
\ref{Wei86}. Contrastingly, the dynamics following from the time-asymmetric 
action is well defined in terms of second-order differential equations of 
motion. Thus this action can be considered as the classical background model 
of relativistic two-quark quantum dynamics. Following$^{\ref{D-T93}}$ we 
reformulate this model into the Lagrangian formalism. Then we transit to the 
Hamiltonian formalism, and integrate the two-body problem in quadratures.

	The time-asymmetric analogue of Rivacoba-Weiss model is the
simplest version of relativistic confinement model. It can be appropriate
for the classical description of light mesons for which the
confinement interaction dominates. To include into consideration also heavy
mesons one can modify the present model by adding to the action the
vector-type interaction term from the time--asymmetric version of the
Wheeler-Feynman electrodynamics. This corresponds to the taking account
of Abelian solution to the standard Yang--Mills equations (i.e., the
classical analogue of one-loop correction in QCD). In the nonrelativistic
limit this mixture leads to the well known Coulomb plus linear potential.
The modified model becomes appreciably cumbersome but still remains 
solvable.

	Here we do not propose a quantum version of the present model.
Instead, we make some estimates of the Regge trajectory from classical
and semiclassical considerations and obtain a physically reasonable result.

	The paper is organized as follows. In Section 2 we obtain the 
Abelian potentials of moving pointlike sources from the standard and
nonstandard theories of Yang-Mills field. The formers are the Lienard-%
Wiechert potentials and their causal modifications while the latters turn
out to be the modifications of potentials proposed by Weiss. They are
obtained with the Green's functions found in Appendix A. In Section 3
we present equations of particle motion following from the standard and
nonstandard theories, and construct corresponding time-symmetric and
time-asymmetric Fokker-type integrals. The latter is used as the base
of time-asymmetric confinement model. In Section 4 this model is 
reformulated in the framework of the Hamiltonian formalism. Various
special cases of two-body problem are considered in Subsections 4.1--4.3
and Appendix B. Estimates of Regge trajectory are quoted in Section 5.
Section 6 is devoted to general discussion of the model.

\section{Abelian potentials from the standard and nonstandard theories
of Yang-Mills field}

	We shall consider both the standard and nonstandard classical
theories of the Yang-Mills field. The standard theory (ST) is based on the
well known Yang-Mills Lagrangian$^{\ref{S-F88}}$
%
\begin{equation}
{\cal L}_{\rm ST}\ =\ -\frac{1}{16\pi}\langle\s F_{\mu\nu},
\s F^{\mu\nu}\rangle -\langle\s J^\mu,\s A_\mu\rangle.
\end{equation}
The nonstandard theory (NT) proceeds from the effective 
Lagrangian$^{\ref{AAB82},\ref{A-A84}}$
%
\begin{eqnarray}
{\cal L}_{\rm NT}&=&\frac{1}{16\pi\kappa^2}\langle\nabla_\lambda
\s F_{\mu\nu}, \nabla^\lambda\s F^{\mu\nu}\rangle +
\frac{\xi}{24\pi\kappa^2}\langle\s F_\mu^{\ \nu},
[\s F_{\nu\lambda},\s F^{\lambda\mu}]\rangle \nonumber\\
&& -\ \langle\s J^\mu,\s A_\mu\rangle.
\end{eqnarray}
Here the components of the gauge field $\s A_\mu(x)$ ($\mu=\overline{0,3}$)
and the current of sources $\s J^\mu(x)$ take
values in the Lie algebra $\cal G$ of gauge group; $[\s X,\s Y]$,
$\langle\s X,\s Y\rangle$, and $\nabla_\mu\s X\equiv\partial_\mu\s X -
[\s A_\mu,\s X]$ are the Lie brackets, the Killing--Cartan metrics, and the
covariant derivative, respectively, defined for any $\s X,\s Y\in{\cal G}$;
$\s F_{\mu\nu}\equiv\partial_\mu\s A_\nu - \partial_\nu\s A_\mu - [\s A_\mu,
\s A_\nu]$ is the tension tensor; $\kappa$ is some parameter of the dimension
of inverse length, and $\xi$ is meant here as an arbitrary dimensionless
parameter. We suppose that the gauge group is the semi-simple compact group,
so that the Killing--Cartan metrics is nondegenerate and positively defined.
In the real matrix representation it can be presented in the form
\begin{displaymath}
\langle\s X,\s Y\rangle = -\frac{1}{N_R}{\rm tr}(\s X\s Y),
\end{displaymath}
where the number $N_R$ depends on the representation chosen. Greece
indices move due to the metrics $\eta_{\mu\nu}$ of the Minkowski space--time
$\IM_4$ which is chosen timelike, i.e., $\|\eta_{\mu\nu}\|=
{\rm diag}(+,-,-,-)$.

	Field equations following from the Lagrangians (1) and (2) are
%
\begin{equation}
\nabla_\nu\s F^{\nu\mu} = 4\pi\s J^\mu,
\end{equation}
and
%
\begin{equation}
\left\{2\nabla_\nu\nabla^2 - (1+\xi)\nabla_\lambda\nabla_\nu\nabla^\lambda
+ \xi\nabla^2\nabla_\nu\right\}\s F^{\nu\mu} = 4\pi\kappa^2\s J^\mu,
\end{equation}
respectively. Both of them are compatible provided the current $\s J^\mu$
is covariantly conserved,
%
\begin{equation}
\nabla_\mu\s J^\mu = 0.
\end{equation}

	In the present paper we are interested in the relativistic system of
$N$ pointlike charged particles interacting via the Yang--Mills field. The
current $\s J^\mu$ corresponding to this system is$^{\ref{Won70}}$
%
\begin{equation}
\s J^\mu(x) \equiv \sum_a \s J^\mu_a(x) =
\sum_a \inta d\tau_a \s Q_a\dot z_a^\mu \delta(x-z_a).
\end{equation}
Here $z^{\mu}_a(\tau_a)$ $(\mu= \overline{0,3},\ a=\overline{1,N})$
are the space-time coordinates of $a$th particle world line in $\IM_4$
parametrized by an arbitrary evolution parameter $\tau_a$,
$\dot z^{\mu}_a(\tau_a) \equiv dz^{\mu}_a/d\tau_a$, and $\s Q_a(\tau_a)$
is the charge of $a$th particle. Substituting (6) into (5) one obtains
the Wong equations$^{\ref{Won70}}$ determining the evolution of charges,
%
\begin{equation}
\dot{\s Q}_a = \dot z^{\mu}_a [\s A_\mu(z_a),\s Q_a], \qquad
a=\overline{1,N}.
\end{equation}

	The total action corresponding to field + particle system can be
written down as follows:
%
\begin{equation}
I = \int\! d^4x\,{\cal L} - \sum_a m_a \inta d\tau_a\lr a,
\end{equation}
where ${\cal L}$ is ${\cal L}_{\rm ST}$ or ${\cal L}_{\rm NT}$, and
$m_a$ is the rest mass of $a$th particle.
	The variation of the action (8) over $\s A_\mu$ yields the
field equations (3) or (4). Varying this action with respect to particle
positions $z^\mu_a$ and taking account of (7) one can obtain the following
equations of particle motion:
%
\begin{equation}
\frac{d}{d\tau_a}\frac{m\dot z_{a\mu}}{\lr a} =
\langle\s Q_a,\s F_{\mu\nu}(z_a)\rangle\dot z^\mu_a.
\end{equation}

	In order to determine motion of particles it is necessary to solve
the total set of linked equations, namely, the field equations [(3) or (4)],
the Wong equations (7), and the equations of motion (9).

	We intend to formulate a particle dynamics in the scope of
RDIT. For this purpose one should eliminate field variables $\s A_\mu(x)$
in favour of their expressions in terms of particle positions $z^\mu_a$ and,
possibly, charges $\s Q_a$. In other words, it is necessary to find a
solution to field equations. But this task is very complicated because
of nonlinearity of the problem.

	Here we limit ourselves by search of Abelian solutions to field
equations. Let us suppose that
%
\begin{equation}
\s A_\mu(x) = \s n A_\mu(x),\qquad
\s J^\mu(x) = \s n J^\mu(x),\qquad
\s Q_a(\tau_a) = \s n Q_a(\tau_a)
\end{equation}
etc., where $\s n$ is a unit constant vector in ${\cal G}$. In this
case all Lie-bracketed expressions vanish, in particular,
%
\begin{equation}
\s F_{\mu\nu}(x) = \s n F_{\mu\nu}(x),\qquad
F_{\mu\nu}(x) \equiv \partial_\mu A_\nu(x) - \partial_\nu A_\mu(x),
\end{equation}
and $\nabla_\mu$ reduces to $\partial_\mu$. The Wong equations (7)
yield
%
\begin{equation}
\dot Q_a = 0\qquad\Longrightarrow\qquad
Q_a = q_a = const.
\end{equation}
Then the field equations (3) and (4) reduce to
%
\begin{equation}
\partial_\nu F^{\nu\mu} = 4\pi J^\mu,
\end{equation}
and
%
\begin{equation}
\partial_\nu\Box F^{\nu\mu} = 4\pi\kappa^2 J^\mu,
\end{equation}
respectively, where the current
%
\begin{equation}
J^\mu(x) \equiv \sum_a J^\mu_a(x) =
\sum_a q_a\inta d\tau_a \dot z_a^\mu \delta(x-z_a)
\end{equation}
is conserved identically, i.e., $\partial_\mu J^\mu\equiv 0$. Due to
this fact both the equations (13) and (14) are gauge invariant with
respect to the one-parametric (compact) group of residual symmetry.

	At this point we have come to the linear field equations which
can be solved by means of the Green's function method. In the standard case
we deal exactly with the electromagnetic problem. Using the Lorentz gauge
fixing condition,
%
\begin{equation}
\partial_\mu A^\mu = 0,
\end{equation}
we reduce the equation (13) to d'Alembert equation,
%
\begin{equation}
\Box A^\mu = 4\pi J^\mu,
\end{equation}
and immediately obtain its solution,
%
\begin{equation}
A^\mu = D_\eta*J^\mu,
\end{equation}
where $*$ denotes the convolution, and
%
\begin{equation}
D_\eta(x) = (1+\eta\,{\rm sgn}\,x^0)\delta(x^2),
\end{equation}
is one of the retarded ($\eta=+1$), advanced ($\eta=-1$), or time-symmetric
($\eta=0$) Green's functions of d'Alembert equation.

	Let us consider the equation (14) of the nonstandard theory.
Using the Lorentz condition (16) one reduces it to the following equation:
%
\begin{equation}
\Box^2 A^\mu = 4\pi\kappa^2 J^\mu,
\end{equation}
which is of 4th order. In Appendix A the corresponding retarded, advanced,
and time-symmetric Green's functions are calculated. They are:
%
\begin{equation}
E_\eta(x) = \qu\kappa^2(1+\eta\,{\rm sgn}\,x^0)\Theta(x^2).
\end{equation}
Thus the solution to (20) reads as (18), but with $E_\eta$ instead of
$D_\eta$.

	Actually, the linearity of equations (17) and (20) allows
solutions of more general structure,
%
\begin{equation}
A^\mu = \sum_a A^\mu_a = \sum_a G_{\eta_a}*J^\mu_a,
\end{equation}
where $G_{\eta_a}=D_{\eta_a}$ for ST, and $G_{\eta_a}=E_{\eta_a}$ for NT.
Here $\eta_a$ take values +1, --1, or 0, each own for different particles.

	In an explicit form the solutions (22) can be written down as
follows:
%
\begin{equation}
A^\mu(x) = \sum_a A^\mu_a(x) =
\sum_a q_a\inta d\tau_a \dot z_a^\mu G_{\eta_a}(x-z_a),
\end{equation}
where the quantity $A^\mu_a(x)$ represents the relativistic potential
created by $a$th particle. In both the ST-- and NT--cases each particle
potential (as well as the total sum (23)) satisfies the Lorentz condition
(16).

Up to the numerical factor, the only difference between (19) and (21) is
that the function $\delta(x^2)$ is replaced by $\Theta(x^2)$. This
substitution was guessed by Weiss in Ref. \ref{Wei86} where the time-symmetric
potential (in our case, Eqs (23) with $G_{\eta_a}=E_0,\ a=\overline{1,N}$)
has been proposed for the model of the action-at-a-distance linear
confinement.

\section{Equations of motion and Fokker-type action integrals}

	Now the equations of particle motion can be obtained
in a closed form by substitution of the relativistic potentials (23)
and the constant charges (12) into the right-hand side (r.h.s.) of (9).
In the standard case this procedure leads to an appearance of divergent
self-action terms which can be regularized in usual way.$^{\ref{dGS72}}$
The resulting equations of motion can be presented in the form:
%
\begin{equation}
\frac{d}{d\tau_a}\frac{m\dot z_{a\mu}}{\lr a} =
q_a \sum_{b\not=a}F_{ab\mu\nu}\dot z^\nu_a + R_{a\mu},
\end{equation}
where
%
\begin{equation}
F_{ab\mu\nu} = 2q_b\inta d\tau_b (1+\eta_b\,{\rm sgn}\,z_{ab}^0)
\delta^{\prime}(z_{ab}^2)\left\{z_{ab\mu}\dot z_{b\nu} -
z_{ab\nu}\dot z_{b\mu}\right\},
\end{equation}
$z_{ab}\equiv z_a - z_b$, and
%
\begin{equation}
R_{a\mu} = \frac{2}{3}\eta_a q^2_a\left\{\delta_\mu^\nu -
\frac{\dot z_{a\mu}\dot z_a^\nu}{\dot z_a^2}\right\}
\frac{d}{d\tau_a}\frac{1}{\lr a}\frac{d}{d\tau_a}\frac{\dot z_{a\nu}}{\lr a},
\end{equation}
The self-action terms $R_{a\mu}$ correspond to radiation reaction. They
disappear if fields generated by particles are time-symmetric (i.e., if
$\eta_a=0$).

	In the nonstandard case no divergences and self-action
terms arise. Thus the equations of motion are calculated immediately.
They are described by (24) with $R_{a\mu} = 0$ and
%
\begin{equation}
F_{ab\mu\nu} = \ha\kappa^2q_b\inta d\tau_b (1+\eta_b\,{\rm sgn}\,z_{ab}^0)
\delta(z_{ab}^2)\left\{z_{ab\mu}\dot z_{b\nu} -
z_{ab\nu}\dot z_{b\mu}\right\}.
\end{equation}

	We have obtained the closed set of equations of particle motion
which are not obvious to be directly deducible from the variacion
principle. Below we construct the relevant Fokker-type version of
the theory and examine it consistency with the equations obtained above.

	The purpose is to eliminate field variables from the total
action (8). Using (10)--(12) in (1), (2), and then in (8), one obtains
the action
%
\begin{equation}
I = I_{\rm free} + I_{\rm int} + I_{\rm field},
\end{equation}
where
%
\begin{equation}
I_{\rm free} = - \sum_a m_a \inta d\tau_a\lr a,
\end{equation}
%
%
\begin{equation}
I_{\rm int} = - \inta d^4 x\, J^\mu A_\mu,
\end{equation}
are the same for ST and NT while $I_{\rm field}$ is different:
%
\begin{equation}
I_{\rm field} = - \frac{1}{16\pi}\inta d^4 x\, F^{\mu\nu}F_{\mu\nu},
\end{equation}
for ST, and
%
\begin{equation}
I_{\rm field} =  \frac{1}{16\pi\kappa^2}\inta d^4 x\,
(\partial^\lambda F^{\mu\nu})(\partial_\lambda F_{\mu\nu}),
\end{equation}
for NT. The term $I_{\rm field}$ can be transformed to the form
%
\begin{equation}
I_{\rm field} = \frac{1}{8\pi}\inta d^4 x\,
A_\mu\partial_\nu H^{\nu\mu} +
\left({\rm surface}\atop {\rm terms}\right),
\end{equation}
where $H^{\nu\mu}=F^{\nu\mu}$ for ST, and $H^{\nu\mu}=
\Box F^{\nu\mu}/\kappa^2$ for NT. Taking into account the field equations
(13) and (14) in r.h.s.
of (33) and omitting surface terms, we obtain
%
\begin{equation}
I = I_{\rm free} + \ha I_{\rm int}.
\end{equation}
Now substituting the current (15) and the potential (23) into
(30), one can present the second term in r.h.s. of (34) in the following
form:
%
\begin{equation}
\ha I_{\rm int} = \sumab I_{ab} + \ha\sum_a I_{aa},
\end{equation}
where
%
\begin{equation}
I_{ab} =
-q_a q_b\intab d\tau_a d\tau_b\,\dot z_a\!\cdot\!\dot z_b\, 
G_{\eta_{ba}}(z_{ab}),
\end{equation}
and $\eta_{ba}\equiv \ha(\eta_b -\eta_a)$.
In the ST--case the self-action term $I_{aa}$ diverges. It can be regularized
and unified with $a$th term of $I_{\rm free}$. In the NT--case
this term vanishes. Thus in the both cases the resulting interaction term
$\ha I_{\rm int}$ has the form:
%
\begin{equation}
\ha I_{\rm int} = - \sumab
q_a q_b\intab d\tau_a d\tau_b\,\dot z_a\!\cdot\!\dot z_b\, 
G_{\eta_{ba}}(z_{ab}),
\end{equation}
where $G_{\eta_{ba}}=D_{\eta_{ba}}$ for ST, and $G_{\eta_{ba}}=
E_{\eta_{ba}}$ for NT.

	In the case of NT each constituent (36) by means of integration
via parts (see Ref. \ref{Kat69} for such a technique) can be transformed to
the following form (here we omit all unessential constant factors):
%
\begin{eqnarray}
\lefteqn{
\int\limits_{-\infty}^{+\infty}
\int\limits_{-\infty}^{+\infty}
d\tau_a d\tau_b\, \dot z_a\!\cdot\!\dot z_b\,
(1+\eta_{ba}\,{\rm sgn}\,z_{ab}^0)\Theta(z_{ab}^2)} \nonumber \\
&=&-2\int\limits_{-\infty}^{+\infty}
\int\limits_{-\infty}^{+\infty}
d\tau_a d\tau_b\,(z_{ab}\cdot\dot z_a)(z_{ab}\cdot\dot z_b)
(1+\eta_{ba}\,{\rm sgn}\,z_{ab}^0)\delta(z_{ab}^2) \nonumber \\
&& -\frac{1}{2}\Biggl[\Biggl[
(1+\eta_{ba}\,{\rm sgn}\,z_{ab}^0)\Theta(z_{ab}^2)z_{ab}^2
\Biggr]_{\tau_a=-\infty}^{\tau_a=+\infty}
\Biggr]_{\tau_b=-\infty}^{\tau_b=+\infty}.
\end{eqnarray}
The second term in r.h.s. of (38) is divergent, but it does
not contribute in equations of motion and can be omitted. Then the
interaction term (37) for NT can be put in the equivalent form,
%
\begin{equation}
\ha I_{\rm int} =
\frac{\kappa^2}{2} \sumab q_a q_b\intab d\tau_a d\tau_b\,
(z_{ab}\cdot\dot z_a)(z_{ab}\cdot\dot z_b) D_{\eta_{ba}}(z_{ab}).
\end{equation}

	Fokker--type equations of motion following from this action differ
from those (24) directly obtained from the field theory. Firstly, they do
not reproduce the self-action terms $R_{a\mu}$ which, in general, are
present in r.h.s. of equations (24) for ST. In this paper we suppose
that these terms can be neglected since in QCD a radiation is suppressed by
confinement. Secondly, the sign factors $\eta_b$ in the expressions (25)
and (27) for $F_{ab\mu\nu}$ are replaced by $\eta_{ba}$. This changes the
causal structure of pair particle interactions. Namely, while equations
(24) correspond to retarded, advanced, or time-symmetric fields generated
by $b$th particles (and sensed by $a$th particle) for $\eta_b$ = +1, --1,
or 0, respectively, in the Fokker--type equations the causality of
interactions is its own for different pairs of particles.

	There are only two cases in which the direct interaction can be
treated as a field-type one. The first case corresponds to the time-%
symmetric interaction, for which $\eta_a = \eta_{ba} = 0$, $a,b =
\overline{1,N}$. For ST the action (34), (29), (37) in this case coincides
with the Wheeler--Feynman action of time-symmetric 
electrodynamics.$^{\ref{Whe49}}$ For NT it corresponds to the action-at-a 
distance confinement model in the form by Weiss.$^{\ref{Wei86}}$ The 
Rivacoba's form of this action integral$^{\ref{Riv84}}$ follows from (39).

	The second case which is tractable in terms of field interaction
realizes only for two-particle systems. It corresponds to the choice $\eta_2
= - \eta_1 = \eta_{21} \equiv\eta=\pm 1$. For ST this is the case of the
time-asymmetric electromagnetic interaction proposed by Staruszkiewicz,
Rudd and Hill,$^{\ref{Sta70}}$ and studied in more detail
in Ref. \ref{Kun74}.
For NT the corresponding time-asymmetric Fokker-type action can be taken
as the classical base for relativistic confinement model.

\section{Time--asymmetric model with confinement interaction}

	In this section we consider the two-particle model available for
the classical description of mesons. It is based on the time-asymmetric
Fokker-type action which combines interaction terms (37) from ST and
(39) from NT. Since mesons are chargeless systems, we put $q_1 =
-q_2 \equiv q$. Then the time-asymmetric Fokker-type action has the
form:
%
\begin{eqnarray}
I& =& -\ \sum\limits_{a=1}^{2} m_a\inta d\tau_a \lr a +
\intab d\tau_1 d\tau_2 D_\eta(z_{12})\times
\nonumber \\
&&\times\left\{\alpha\,\dot z_1\!\cdot\!\dot z_2 -
\beta (z_{12}\cdot\dot z_1)(z_{12}\cdot\dot z_2)\right\},
\end{eqnarray}
where $\alpha\equiv q^2$, $\beta\equiv \ha q^2\kappa^2$, and $\eta=\pm 1$.
In the nonrelativistic limit this action leads to the well known interquark
potential $U = - \alpha/r + \beta r$.

	Integrating the second term of the action (4) once, we reduce the
latter to a single-time form.$^{\ref{D-T93}}$ Thus we obtain the
description of our model in the framework of a manifestly covariant
Lagrangian formalism with the Lagrangian function
%
\begin{equation}
L= \theta F(\sigma_1,\sigma_2,\delta),
\end{equation}
where $\theta \equiv \eta \dot y \!\cdot\! z > 0$, ~$z \equiv z_1 - z_2$,
~$y \equiv (z_1 + z_2)/2$,
~$\sigma_a \equiv \lr{a}/\theta>0$, ~$\delta \equiv \dot z_1\cdot\dot z_2/
\theta^2>0$, and with the holonomic constraint $z^2=0$, $\eta z^0>0$.
All variables in (41) depend on an arbitrary common evolution parameter
$\tau$. In our case the function $F$ has the form:
%
\begin{equation}
F \equiv \sum^2_{a=1} m_a \sigma_a - \alpha\delta + \beta.
\end{equation}
We note that quantities $\theta,\sigma_a,\delta$ in r.h.s. of (41) and (42)
are well defined and positive if particle world lines are timelike.

	The transition to the manifestly covariant Hamiltonian description
with constraints leads to the mass-shell constraint which determines the
dynamics of the model and has the following form:$^{\ref{D-T93}}$
%
\begin{equation}
\phi(P^2,~\upsilon^2, ~P \cdot z, ~\upsilon \cdot z) \equiv \phi_{\rm free}
 + \phi_{\rm int}=0.
\end{equation}
Here $\upsilon_{\mu}\equiv w_{\mu}-z_{\mu}\,P\!\cdot\!w/P\!\cdot\!z;~P_{\mu}$
and $w_{\mu}$ are canonical momenta conjugated to $y^{\mu}$ and $z^{\mu}$,
respectively; the function
%
\begin{equation}
\phi_{\rm free}= \qu P^2 - \ha (m^2_1 + m_2^2) + (m_1^2 - m_2^2)
\frac{v\!\cdot\!z}{P\!\cdot\!z} + v^2
\end{equation}
corresponds to the free-particle system,
%
\begin{eqnarray}
\phi _{\rm int} & = & \frac{\alpha
(P^{2}-m_{1}^{2}- m_{2}^{2})}{\eta P\!\cdot\!z}
 +\,\frac{\alpha^2}{\eta P\!\cdot\!z}\sum^2_{a=1}
\frac {m_a^2}{b_a+\alpha}  \nonumber\\
&&-\,2\beta\left(\frac{b_1 b_2}{\eta P\!\cdot\!z} + \alpha\right)
\end{eqnarray}
describes the interaction, and
%
\begin{equation}
b_a \equiv \eta \left(\ha P\!\cdot\!z + \na v\!\cdot\!z \right),~~~~
a=1,2,~~~~\bar a \equiv 3-a.
\end{equation}

	We note, that the quantities $\sigma_a$ are related to
canonical variables by the equations:
%
\begin{equation}
\sigma_a = m_a/(b_a+\alpha),~~~~~~~~~~~~~~~a=1,2.
\end{equation}
Since $\sigma_a$ must be positive, the following conditions arise:
%
\begin{equation}
b_a+\alpha>0,~~~~~~~~~~~~~~~a=1,2.
\end{equation}
They restrict the whole phase space to a physical domain in which the
Hamiltonian description is equivalent to the Lagrangian one.

	In order to study the dynamics of the present model it is
convenient, following Ref. \ref{D-T93}, to transit from the manifestly
covariant to three-dimensional Hamiltonian description in the framework of
the Bakamjian-Thomas model.$^{\ref{B-T53},\ref{D-K92}}$ Within this 
description ten generators of the Poincar\'e group $P_{\mu}$,
$J_{\mu\nu}$ as well as the covariant particle positions $z_a^{\mu}$ are the
functions of canonical variables ${\bf Q,~P, ~r, ~k}$. The only arbitrary
function appearing in expressions for canonical generators is the
total mass $| P |=  M({\bf r, k)}$ of the system which determines
its internal dynamics. For the time-asymmetric models this function is
defined by the mass-shell equation$^{\ref{D-T93}}$ which can be derived
from the mass-shell constraint via the following substitution of arguments
on the l.-h.s. of (43):
%
\begin{equation}
P^2\!\to M^2,~~v^2\!\to -{\bf k}^2,~~P\!\cdot\!z\!\to\!\eta
M r,~~ v\!\cdot\!z\!\to - {\bf k}\!\cdot\!{\bf r};
\end{equation}
here $r \equiv$ $| {\bf r}|$.

	Due to the Poincar\'e-invariance of the description it is sufficient
to choose the centre-of-mass (CM) reference frame in which
${\bf P}\!=\!{\bf 0},\ {\bf Q}\!=\!{\bf 0}$. Accordingly,  $P_0 = M$,
$J_{0i}=0~ (i=1,2,3)$, and the components $S_i \equiv \frac{1}{2}
\varepsilon_i^{\ jk}J_{jk}$ form a 3-vector of the total spin of the system
(internal angular momentum) $\B S=\B r\times\B k$ which is the integral of
motion. At this point the problem is reduced to the rotation invariant
problem of some effective single particle; such a problem is integrable in
terms of polar coordinates,
%
\begin{equation}
{\bf r} = r{\bf e}_r,~~~~{\bf k} = k_r{\bf e}_r + S{\bf e}_{\varphi}/r.
\end{equation}
Here $S \equiv | {\bf S} |$; the unit vectors ${\bf e}_r$,
~${\bf e_{\varphi}}$ are orthogonal to ${\bf S}$, they form together
with ${\bf S}$ a right-oriented triplet and can be decomposed in terms of
Cartesian unit vectors ${\bf i,~j}$:
%
\begin{equation}
{\bf e}_r = {\bf i} \cos \varphi + {\bf j} \sin \varphi,~~~~
{\bf e}_{\varphi} = - {\bf i} \sin \varphi + {\bf j} \cos \varphi,
\end{equation}
where $\varphi$ is the polar angle.

	The corresponding quadratures read:
%
\begin{eqnarray}
t - t_0 = \inta dr\ \partial k_r(r,M,S)/\partial M,&& \\
\varphi - \varphi_0 = - \inta dr\ \partial k_r(r,M,S)/\partial S,&&
\end{eqnarray}
where $t=\frac{1}{2}(z_1^0 + z_2^0)_{{\rm CM}}$ is the fixed evolution
parameter (unlike the undetermined one $\tau$), and the radial momentum
$k_r$, being the function of $r,~M,~S$, is defined by the mass-shell
equation written down in
terms of these variables,
%
\begin{eqnarray}
\lefteqn{
\phi\left(M^2,\ - \B k^2,\ \eta Mr,\ -\B k\cdot\B r\right)}  \nonumber\\
& \equiv &
\phi\left(M^2,\ - k_r^2 - \frac{S^2}{r^2},\
\eta Mr,\ -k_r r\right) = 0.
\end{eqnarray}

	The solution of the problem given in terms of canonical variables
enables to obtain particle world lines in the Minkowski space using the
following formulae:$^{\ref{D-T93}}$
%
\begin{equation}
z_a^0 = t + \ha\na \eta r,
\end{equation}
\begin{equation}
{\bf z}_a = \ha\na\B r +
\eta r\frac{\B k}{M} \equiv
\left(\frac{1}{2}(-)^{\bar a} +
\eta \frac{k_r}{M} \right) r
\B e_r + \eta \frac{S}{M} \B e_\varphi.
\end{equation}
Especially, the vector $\B z = \B z_1 - \B z_2 = \B r$
characterizes the relative motion of particles.

\subsection{Purely confinement model}

	Hereafter we restrict ourselves to the system of equal
rest masses, $m_1 = m_2 \equiv m$. The case $\alpha=0$ corresponds to
purely confinement interaction. The mass-shell equation in this case
reads:
%
\begin{equation}
\frac{S^2}{r^2} + m^2 - \left(1 - 2\frac{\beta r}{M}\right)
\left(\qu M^2 - k_r^2\right) = 0.
\end{equation}
It easy to obtain from (57) the expression for $k_r(r,M,S)$,
%
\begin{equation}
k_r = \epsilon \sqrt{f(r,M,S)}, \qquad \epsilon=\pm1,
\end{equation}
%
%
\begin{equation}
f(r,M,S) = \qu M^2 - \frac{m^2 + S^2/r^2}{1 - 2\beta r/M} \ge 0.
\end{equation}
Besides, we must take into account the condition:
%
\begin{equation}
\qu M^2 - k_r^2 > 0
\end{equation}
which follows from (48). Then from (58)--(60) we obtain the restriction:
%
\begin{equation}
0<r<\ha M/\beta.
\end{equation}

	The quadratures (52), (53) with (58), (59) can be reduced
to the elliptic integrals. Here we omit their expressions. The integration
is spread over the domain of possible motions (DPM) which is determined
by the conditions (59) and (61). In the case $S>0$ DPM consists of
the connected interval $r_1\le r \le r_2$, where $r_1,\ r_2$ are
positive roots of the equation $f(r,M,S) = 0$. The latter can be presented
as the reduced cubic equation with respect to $1/r$:
%
\begin{equation}
\frac{1}{r^3} - \frac{M^2}{4S^2}\left(1 - \frac{4m^2}{M^2}\right)\frac{1}{r}
+ \frac{M\beta}{2S^2} = 0.
\end{equation}
It has two real positive solutions provided the following condition holds:
%
\begin{equation}
M\ge M_c(S),	
\end{equation}
where the function $M_c(S)$ is defined in the implicit form
%
\begin{equation}
S = \frac{M_c^2}{6\sqrt{3}\beta}\left(1 -
\frac{4m^2}{M_c^2}\right)^{3/2}, \qquad M_c\ge 2m.
\end{equation}
The equality in (63) corresponds to the case $r_1=r_2\equiv r_c$ of circular
particle orbits with the distance between particles
%
\begin{equation}
r_c = \frac{M_c}{3\beta}\left(1 -
\frac{4m^2}{M_c^2}\right)
\end{equation}
satisfying the set of equations:
%
\begin{equation}
f(r_c, M_c, S) = 0, \qquad 
\partial f(r_c, M_c, S) / \partial r_c = 0.
\end{equation}

	In the limit $S\to 0$ the quadrature (53) yields $\varphi=\varphi_0$,
and a particle motion becomes one-dimensional (i.e., in the two-dimensional
space-time $\IM_2$ parametrized with $x^0$ and, say, $x^1$). 
Besides, $r_1\to0$. Thus DPM becomes $0<r\le r_2$. The 
point $r=0$ corresponds to particle collision. This point is not singular
for the quadrature (52) and particle coordinates (55), (56). Thus the motion 
of particles can be smoothly continued as if they pass through one another.

\subsection{General model, $S>0$}

	Let us consider the general case $\alpha>0$, $\beta>0$.
The corresponding mass-shell equation can be written down as follows:
%
\begin{equation}
\frac{S^2}{r^2}\Delta -
\left(\Delta - \frac{\alpha^2}{r^2}\right)
\left\{\left(1 - 2\frac{\beta r}{M}\right)\Delta -
2\frac{m^2}{M}\left(\frac{M}{2} + \frac{\alpha}{r}\right)\right\} = 0.
\end{equation}
It is quadratic equation with respect to
%
\begin{equation}
\Delta \equiv  \left(\frac{M}{2} + \frac{\alpha}{r}\right)^2 - k_r^2 > 0,
\end{equation}
where $\Delta$ must be positive because of the conditions (48). As to $k_r$
the equation (67) is biquadratic. Its solution can be presented in the
following form:
%
\begin{equation}
k_r = \epsilon \sqrt{f_\pm(r,M,S)}, \qquad \epsilon=\pm1,
\end{equation}
where
%
\begin{eqnarray}
f_\pm(r,M,S)&=&\qu M^2 + M\frac{\alpha}{r}-
\frac{h_\pm(r,M,S)}{1 - 2\beta r/M},	\\
h_\pm(r,M,S)&=&g(r,M,S) \mp\sqrt{d(r,M,S)} \\
d(r,M,S)&=&
g^2(r,M,S) + \left(1 - 2\frac{\beta r}{M}\right)
\frac{\alpha^2 S^2}{r^4},	\\
g(r,M,S)&=& \frac{m^2}{M}\left(\frac{M}{2} + \frac{\alpha}{r}\right) -
\left(1 - 2\frac{\beta r}{M}\right)\frac{\alpha^2}{2r^2} + \frac{S^2}{2r^2}.
\end{eqnarray}
Among two solutions $f_\pm$ for $k_r^2$ we choose that one which is smooth
in DPM and reduces to $f$ (59) in the limit $\alpha\to0$.

	 DPM is analyzed in Appendix B. In the case $S>0$ we have
$d(r,M,S)>0$, $r>0$. Thus both the functions $f_\pm(r,M,S)$ are smooth
provided $r\ne\ha M/\beta$, and $f_-(r,M,S)$ reduces to $f$ (59) in the
limit $\alpha\to0$. DPM in this case is determined by inequality
$f_-(r,M,S)\ge 0$ provided the condition (68) holds. Similarly to the purely
confinement case we obtain $r_1\le r \le r_2$, where $r_1,\ r_2$ are 
positive roots of the equation:
%
\begin{eqnarray}
\lefteqn{
\frac{S^2}{r^2}
\left(\frac{M}{2} + \frac{\alpha}{r}\right)
-\left(
\qu M^2 + M\frac{\alpha}{r}\right)\times}
\nonumber\\
&&\times\left\{\left(1 - 2\frac{\beta r}{M}\right)
\left(\frac{M}{2} + \frac{\alpha}{r}\right)
-2\frac{m^2}{M}\right\} = 0
\end{eqnarray}
\begin{figure}[h,t]
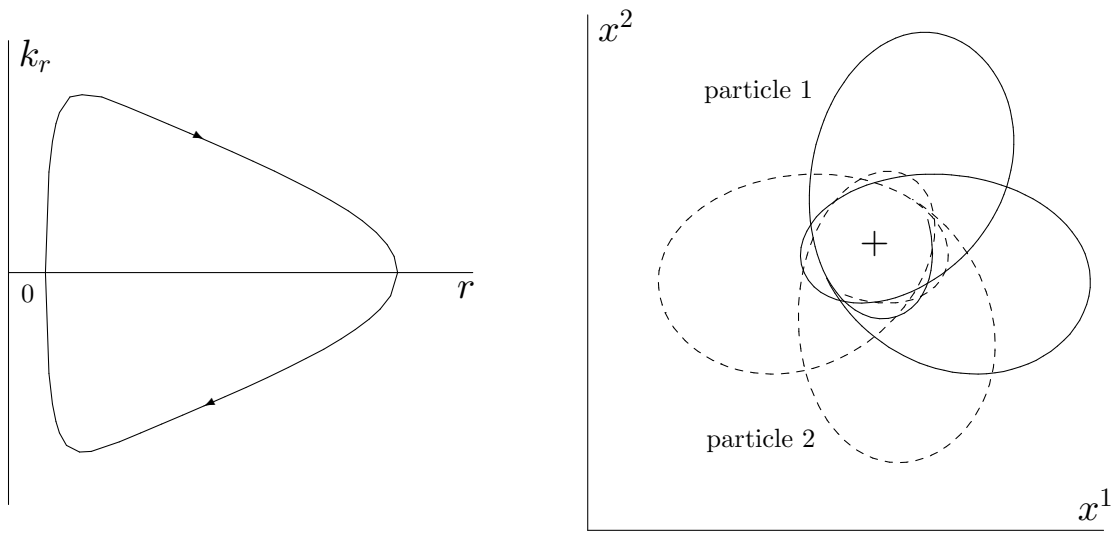

\vspace{2ex}
\input ijmp-f1.pic
\caption{{\normalsize General model, $S>0$. Typical example of phase (left) 
and particle (right) trajectories. Arrows show the direction of evolution; 
+~-- centre of mass.}
}
\end{figure}
It can be reduced to a fourth-order algebraic equation which has two real
positive solutions provided (63), where the function $M_c(S)$
can be presented in a parametric form,
%
\begin{eqnarray}
M_c^2(\lambda)&=&2m^2\frac{\lambda[4+5\lambda+2\lambda^2+\nu(1+\lambda)^2
(4+3\lambda)]}{(1+\lambda)^3}, \nonumber\\
S^2(\lambda)&=&\alpha^2\frac{\lambda(2+\lambda)^2[1+\nu(1+\lambda)^2]}
{4+5\lambda+2\lambda^2+\nu(1+\lambda)^2
(4+3\lambda)}, \nonumber\\
\nu&\equiv&\frac{\alpha\beta}{m^2}, \qquad 0<\lambda<\infty.
\end{eqnarray}

The condition $M=M_c(S)$ corresponds to circular orbits with
$r_c=2\alpha\lambda/M_c$ to be the distance between particles.

	Our attempts to express	the quadratures (52), (53) with (69)--(73) 
in terms of known (elementary and special) functions have not met with
success. Thus we calculated them with a computer.
Nevertheless by means of analytic calculations it can be
shown that particle world lines in $\IM_4$ are timelike and smooth
curves. They represent a bound motion of particles for all values of $M$ 
allowed by (63), (75), and $S>0$. The typical example of phase and particle
trajectories are shown in figure 1.

\subsection{General model, $S=0$}

	In the case $S=0$ we have $d(r,M,0) = g^2(r,M,0)$. Since there 
exists the point $r_0<\alpha/m$ such that $g(r_0,M,0)=0$, the functions 
$f_\pm(r,M,0)$ are not smooth. Moreover, in the domain $r<r_0$ the function 
$f_-(r,M,0)$ has not the proper form in the limit $\alpha\to0$. Thus the 
evolution of particles cannot straightforwardly be continued farther. 

	We point out that the distance $r_0$ at which the smoothness of
$f_\pm(r,M,0)$ violates is smaller than $\alpha/m$. The latter is an analogue
of the classical electron radius. In the case of strong interaction the 
distance $r_0$ and the Compton length of quarks can be commensurable
quantities. Thus the
classical description of particle motion at $r<r_0$ may be important for
the construction of quantum theory. Especially, this is concerned with the
case of S--states. Below we propose the way to continue the particle motion
in the domain $r<r_0$. It leads beyond the rigorous treatment of analytical
mechanics and therefore cannot be a reliable basis of quantum-mechanical 
description. But it will be noted that this method arises naturally from the 
present model itself.

	Let us choose in r.h.s. of (69) the function:
%
\begin{eqnarray}
f_0(r,M)&\equiv&\left\lbrace f_+(r,M,0),\quad r<r_0\atop
f_-(r,M,0),\quad r>r_0\right. \nonumber \\
&=&\left(\frac{M}{2} + \frac{\alpha}{r}\right)
\left\lbrace\frac{M}{2} + \frac{\alpha}{r} - \frac{2m^2/M}{1 -
2\beta r/M}\right\rbrace,
\end{eqnarray}
which is smooth provided $r\ne\ha M/\beta$, and reduces to (59) if 
$\alpha\to0$. DPM in this case is $0<r\le r_2$ while the point $r=0$
is critical: $\Delta\to\infty$, $r\to0$. This means that the equivalence
between the Lagrangian and Hamiltonian formalisms violates. It can be shown 
that at the collision one of particles reaches (but not exceeds) the speed of 
light while another does not. Again, the particle world lines should be
somehow continued farther.
\begin{figure}[p]
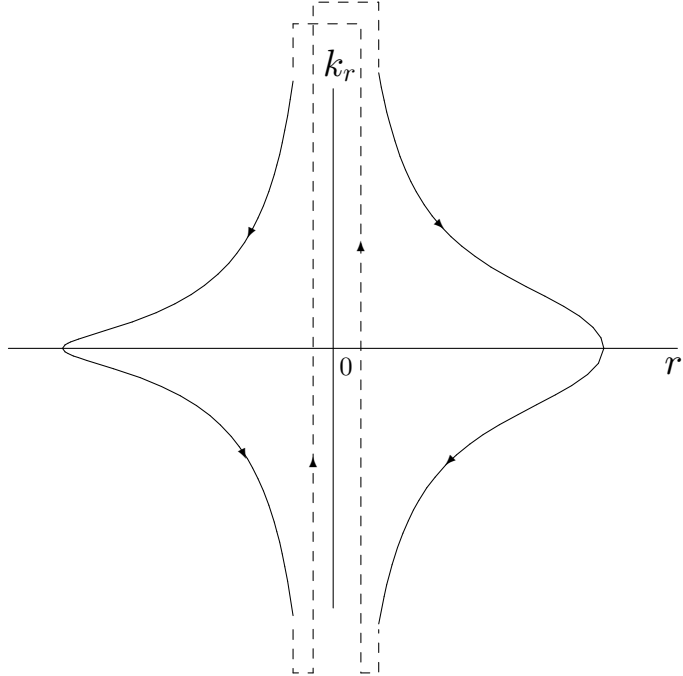

\input ijmp-f2.pic
\vspace{-0.5ex}
\caption{{\normalsize General model, $S=0$. Typical example of phase 
trajectory continued in the non-Lagrangian domain $r<0$. Arrows show the 
direction of evolution.}}
\end{figure}
\begin{figure}[p]
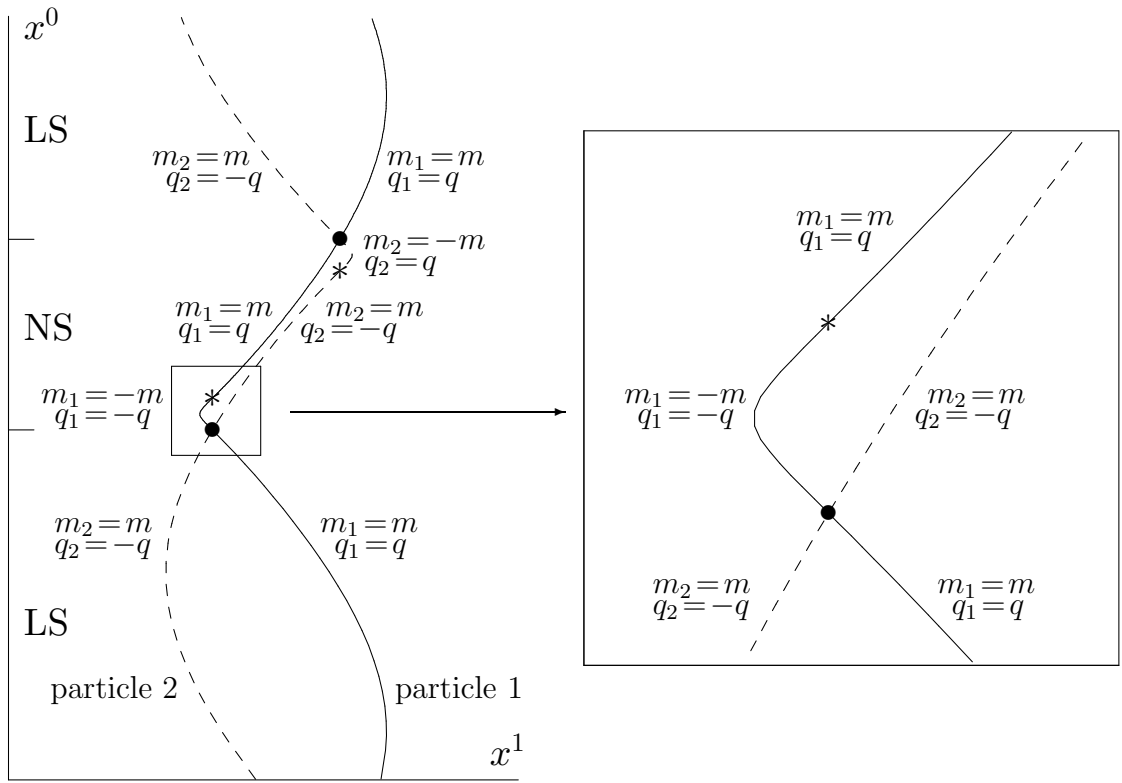

\vspace{1ex}
\input ijmp-f3.pic
\vspace{-0.5ex}
\caption{{\normalsize General model, $S=0$. Typical example of particle 
world lines. LS~-- Lagrangian segments; NS~-- non-Lagrangian segments; 
$\bullet$~-- collision: one of particles reaches the light speed; $\ast$~-- 
particles reach the light speed.}
}
\end{figure}

	The existence of a singular collision point is due to the time-%
asymmetric vector interaction from ST. The confinement interaction does
not change qualitatively the behaviour of particles in the neighbourhood of
collision. Specific features of the time-asymmetric model with attractive
vector interaction in $\IM_2$ have been analyzed in Ref. \ref{Shp98}. 
Following this work, in the framework of Hamiltonian description the 
interesting possibility exists to continue smoothly world lines through the 
collision point. In our terms it is sufficient for this purpose to suppose 
formally that after the collision the variable $r$ becomes negative. Then we 
have $f_0(r,M) \ge 0$, $r\in[-2\alpha/M,0)$, and the motion of particles can 
be continued up to the distance $|r|=2\alpha/M$. At this point which is also
critical both particles reach (but not exceed) the speed of light.
Again, one can smoothly continue world lines up to the next collision e.t.c.
We note that $\Delta<0$, $r\in(-2\alpha/M,0)$. Thus the segments of
world lines obtained as above do not follow from the Lagrangian description.
The resulting world lines combine the Lagrangian and non-Lagrangian segments
separated by the collision points. They describe the bound periodic motion
of particles. The corresponding phase trajectory and world lines are shown
in figure 2 and 3 respectively.
			
	The formal continuation of evolution proposed above permits some
reinterpretation in terms of the Lagrangian description. Expressing
the quantities $\theta$, $\sigma_a$ and $\delta$ in terms of canonical
variables one can examine that some of them have wrong (i.e., negative)
sign if $r<0$, i.e., if particles pass non-Lagrangian segments of world
lines. Equivalently, one can keep $r>0$ changing  signs of some constants
$m_a$, $\alpha$, and $\beta$ in the Lagrangian (41), (42). In such a manner
one can realize that particles move as if each one changes signs of its rest 
mass and charge, $m_a\to-m_a$, $q_a\to-q_a$, once it passes a critical point
with the speed of light. We note that at the collision point one of world
lines is timelike. Thus the mass and charge of this particle remain 
unchanged up to the next critical point. As a result, after having passed
the non-Lagrangian segment each particle returns its proper values of
mass and charge.

\section{Semiclassical estimates of Regge trajectory}

	It is well known that nonrelativistic potential model
with the linear potential leads to the Regge trajectory with the
unsatisfactory asymptote $M \sim S^{2/3}$. Here we do not propose
a quantum version of the present model, but we make the estimates of the
Regge trajectory from what follows.

	Usually the Regge trajectories in the potential models are
calculated in the oscillator approximation.$^{\ref{L-S89}}$ Then the leading
Regge trajectory originates from the classical mechanics: it is close to
the curve of circular motions on the ($M^2,S$)--plane.

	In our case this curve follows from the equation (64) for $\alpha=0$
or from (75) in the general case. The latter has in the ultrarelativistic 
limit $M_c\to\infty$ the desirable linear asymptote:
%
\begin{equation}
M_c^2 \approx 6\sqrt{3}\beta S + 6\left(m^2 - 3\alpha\beta\right).
\end{equation}
It is remarkable that this asymptote is achieved only by taking account
of a relativity.
\begin{figure}[h,t]
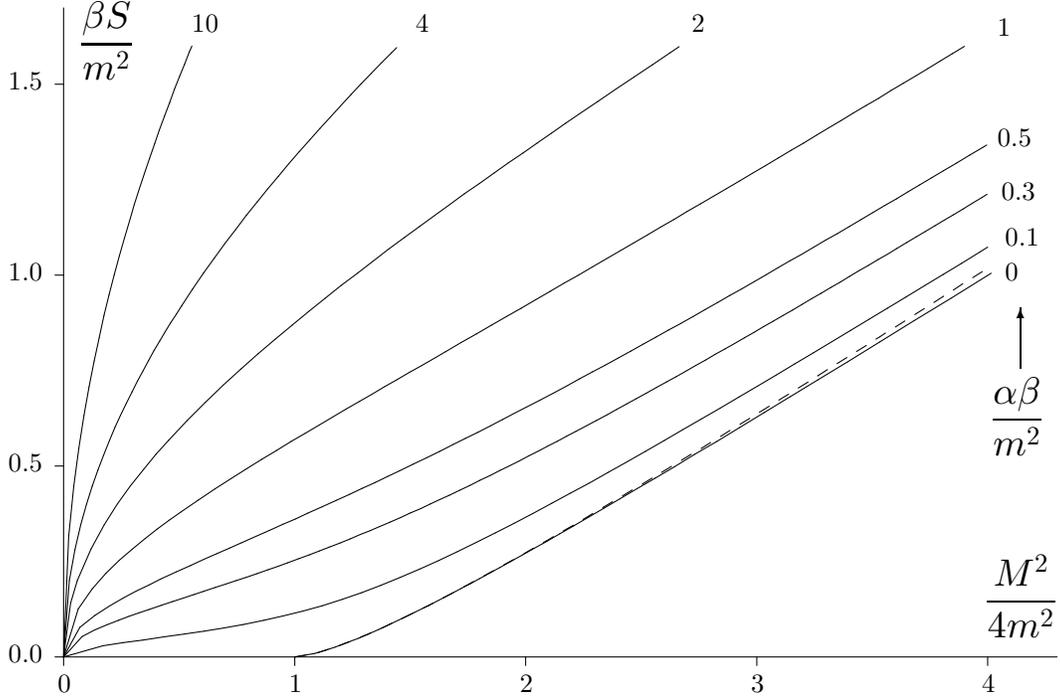

\input ijmp-f4.pic
\caption{{\normalsize Classical Regge trajectories; -------- from general
time-asymmetric model at various rates of parameters; --~--~-- from 
time-symmetric model.}
}
\end{figure}

	Let us compare the classical Regge trajectory (64) of purely 
confinement time-asymmetric model to that which follows from the 
time-symmetric Fokker-type confinement model with the same parameters 
$m_1 = m_2 \equiv m$ and $\beta$. Considering the circular orbit solution, 
given in Ref. \ref{Riv84},\footnote{Such a solution given in Ref. \ref{Wei86} 
seems to contain a miscount.} for large $M$ one can obtain:
%
\begin{equation}
M_c^2 \approx 4(1+\sin\vartheta)\{2\cos\vartheta\,\beta S + m^2\},
\end{equation}
where the angle 
%
\begin{equation}
\vartheta \approx 0.7391 \approx 0.2353\pi
\end{equation}
is the solution of the transcendental equation
%
\begin{equation}
\vartheta = \cos\vartheta.
\end{equation}
The only difference between the asymptotes (77) (with $\alpha=0$, of course) 
and (78)  consists in slightly different numerical factors at the linear 
and constant terms
%
\begin{equation}
\begin{array}{ccc}
{\rm linear\ terms}&\quad&{\rm constant\ terms}\\
6\sqrt{3} \approx 10.3923&\quad&6\\
8\cos\vartheta\,(1+\sin\vartheta) \approx 9.8955&\quad&
4(1+\sin\vartheta) \approx 6.6944
\end{array}
\end{equation}
Moreover, it turns out that by the substitution $\vartheta=\pi/6$ instead of 
(79) into the r.h.s. of (78), the latter reduces to (77) (with $\alpha=0$).
In the nonrelativistic limit both the time-symmetric and time-asymmetric
(purely confinement) models lead to the same relation:
%
\begin{equation}
M_c-2m \approx 3\left(\frac{\beta S}{2\sqrt{m}}\right)^{2/3}
\end{equation}
which is known from the nonrelativistic linear confinement model. Classical
Regge trajectories from the general time-asymmetric model as well as one
from time-symmetric model are shown in figure 4.

	These purely classical results give us the base to consideration
of semiclassical quantization of the model. By analogy with WKB approximation
method we put
%
\begin{equation}
S = \hbar(\ell+\ha), \qquad \ell = 0,1,\dots
\end{equation}
for the quantized internal momentum, and
%
\begin{equation}
\oint k_r dr = 2\pi\hbar(n_r + \ha), \qquad n_r = 0,1,\dots
\end{equation}
for radial excitations; the integral runs over the classical phase
trajectory. 

	In the case of purely confinement model we have:
%
\begin{equation}
\int\limits_{r_1}^{r_2}dr\sqrt{f(r,M,S)} = \pi\hbar(n_r + \ha).
\end{equation}
Using the oscillator approximation$^{\ref{L-S89},\ref{GCO89}}$ we 
expand the functi\-on $f(r,M,S)$ (59) about the circular orbit to first 
nonvanishing orders in $\Delta M\equiv M-M_c$ and $\Delta r\equiv r-r_c$. 
The result is as follows:
%
\begin{equation}
f(r,M,S) \approx a^2(M,S) - b^2(S)(\Delta r)^2,
\end{equation}
where
%
\begin{eqnarray}
a^2(M,S)\equiv\left.\frac{\partial f(r,M,S)}{\partial M}\right|_c
\Delta M = \frac{M_c(M_c^2+2m^2)}{(M_c^2+8m^2)}\Delta M, \\
b^2(S)\equiv-\left.\frac{\partial^2 f(r,M,S)}{2\,\partial r^2}\right|_c
= \frac{27\beta^2M_c^4}{4(M_c^2-4m^2)(M_c^2+8m^2)},
\end{eqnarray}
and the function $M_c(S)$ is defined in (64).
Then the integral in l.h.s. of (85) is easily calculated:
%
\begin{equation}
\int\limits_{-a/b}^{a/b}d(\Delta r)\sqrt{a^2-b^2(\Delta r)^2} = 
\frac{\pi}{2}\frac{a^2}{b}.
\end{equation}
Using (85)--(89) and assuming that $\Delta M$ is small compared to $M_c$
we obtain for $M^2$ the expression:
%
\begin{equation}
M^2 = M_c^2\left\{1 + \frac{6\sqrt{3}\beta\hbar}{M_c^2+2m^2}
\sqrt{\frac{M_c^2+8m^2}{M_c^2-4m^2}}\,(n_r + \ha)\right\}
\end{equation}
which together with (83) and the definition (64) of the function $M_c(S)$ 
describes the leading (for $n_r=0$) and daughter (for $n_r=1,2,\dots$) Regge 
trajectories. 
\begin{figure}[p]
\input ijmp-f5a.pic
\vspace{5ex}
\input ijmp-f5b.pic
\caption{{\normalsize Semiclassical Regge trajectories; 
$m=1.27\,{\rm GeV}$, $\alpha=0.5$, $\beta=0.2\,{\rm GeV}^2$;
a) oscillator approximation; b) numerical solution.}
}
\end{figure}
\begin{figure}[p]
\input ijmp-f6a.pic
\vspace{5ex}
\input ijmp-f6b.pic
\caption{{\normalsize Semiclassical Regge trajectories; $m=0.005\,
{\rm GeV}$, $\alpha=0.8$, $\beta=0.2\,{\rm GeV}^2$;
a) oscillator approximation; b) numerical solution.}
}
\end{figure}

	Similarly, in the case of general model we obtain the Regge
trajectories determined in the implicit form by the equations
%
\begin{equation}
M^2 = M_c^2(\lambda)\left\{1 + 
\mbox{{\small$\left.{\vphantom{\hbar}\mbox{\normalsize$\hbar$}}\atop
{\vphantom{\hbar}\mbox{\normalsize$\overline\alpha$}}\right.$}}
\Phi(\lambda)\,(n_r + \ha)\right\},
\end{equation}
%
%
\begin{equation}
\hspace{-1em}
\Phi(\lambda)=\frac{\sqrt{\left[1\!+\!2\lambda\!+\!\nu(1\!+\!\lambda)^2\right]
\left[3\!+\!(1\!+\!\nu)(1\!+\!\lambda)^2\!+\!3\nu(1\!+\!\lambda)^4\right]}}
{(1 + \lambda)^2[1 + \nu(1 + \lambda)]\sqrt{\lambda(2 + \lambda)}},
\end{equation}
and (75), (83). At large $\ell$ these trajectories reduce to linear ones,
%
\begin{equation}
M_c^2 \approx 6\sqrt{3}\beta\hbar(\ell + n_r + 1)
+ 6\left(m^2 - 3\alpha\beta\right),
\end{equation}
so that the daughters are parallel to leading trajectory. Moreover, states
of unit internal momentum differences form into degenerate towers at a given
mass. This tower structure is of interest for the meson spectroscopy, as it 
is intimated in Ref. \ref{GCO89}. The number of relativistic potential models
based on single-particle wave equations$^{\ref{FKR71},\ref{GCO89},
\ref{HLS94}}$ as well as two-particle models with oscillator 
interaction$^{\ref{Tak79},\ref{Pol89}}$ lead to degeneracy of $\ell+2n_r$ 
type, but not of $\ell+n_r$ type.\footnote{In the models presented in Ref. 
\ref{HLS94} the degeneracy appears if to neglect the spin of quarks.} The 
latter cannot be reproduced by single-particle relativistic models with the 
vector and scalar potentials, as it is shown in Ref. \ref{GCO89}. 

	Figures 5 and 6 present two examples of semiclassical Regge 
trajectories which are characteristic for heavy and light mesons 
respectively. Trajectories in figures 5a and 6a are calculated in the
oscillator approximation which is good for $n_r\ll\ell$. Thus
curved segments of daughters at $n_r\ge\ell$ are not sure. This is evident
by comparison to trajectories of the same case which are shown in figures 5b 
and 6b. They are obtained by the numerical solving of the equation (84) where
the integral in l.h.s. of (84) runs over phase trajectories of general model.
We note that due to (83) phase trajectories in this case correspond to
$S>0$ (see figure 1), and thus they are free of critical points discussed in
Subsection 4.3. It is remarkable that numerical solutions for mass spectrum
is well described by the asymptotic formula (93) even at small $\ell$.
This is especially concerned with the case of light mesons (figure 6b).

\section{Conclusion}

	In the present paper we have traced the relation between the
nonstandard classical Yang-Mills field theory which arises from the 
consideration of QCD in the infrared region$^{\ref{AAB82}}$ and the
classical relativistic two-particle models with confinement interaction
formulated in the framework of Fokker formalism. It is notable that
the use of Abelian potentials following from NT provides the confining
interaction of particles. The time-symmetric (purely confinement) model 
turns out known in the literature$^{\ref{Riv84},\ref{Wei86}}$ where, 
although, it was constructed as a priori action-at-a-distance model. 

	The present time-asymmetric confinement model could be regarded as a
classical relativisation of the primitive quarkonium model.
It has a number of features which are expected for models
of this kind but which usually are not realized together.

	1. The model is a self-consistent relativistic two-particle model.
It allows the Lagrangian and Hamiltonian formulations. The quantities in 
terms of which the model is built have clear physical meaning. In the case
$S>0$ solutions of this model lead to smooth timelike particle world lines.
If $S=0$, the collision critical points occur in which particles reach the 
speed of light. In this case, although, world line can be smoothly continued 
as well. In both cases a particle motion is bound.

	2. Estimates of Regge trajectory from classical mechanics shows that 
it has a proper asymptote while the corresponding nonrelativistic potential 
is linear. This feature is not derivable from nonrelativistic models. 
The parameters of a linear rise following from the time-symmetric and 
time-asymmetric models differ from one another by near 5~\%. One can hope 
that other long-range effects which should follow from the forthcoming study 
of purely retarded, time-symmetric and time-asymmetric confinement 
interactions differ slightly as well. 

	The semiclassical consideration leads to the interesting degenerate
tower structure of meson spectrum which probably exists in nature. 

	3. The interpretation of an interaction in terms of some field theory
is very important in RDIT. Hopefully, the knowledge of field equations and 
corresponding variational principle underlying the model allows to include 
properly into consideration spinning particles and then to construct the 
quantum-mechanical description.

\section*{Acknowledgments}

	The author would like to thank Prof. V. Tretyak and Dr. V. Shpytko
for useful discussions and for a helpful reading of this manuscript.

\section*{Appendix} 
\subsection*{A. Green's functions for field equation of NT.}
\renewcommand{\theequation}{A.\arabic{equation}}
\setcounter{equation}{0}

	Let us consider the equation (20) (here we omit unessential
Greece indices and the factor $\kappa^2$),
%
\begin{equation}
\Box^2 A(x) = 4\pi J(x).
\end{equation}
It can be recast into the set of two d'Alembert equations,
%
\begin{equation}
\Box A(x) = B(x),
\end{equation}
%
%
\begin{equation}
\Box B(x) = 4\pi J(x).
\end{equation}
Solving them yields the following formal expression for $A(x)$:
%
\begin{equation}
A = \frac{1}{4\pi}D_\eta*B = \frac{1}{4\pi}D_\eta*D_{\eta^{\prime}}*J,
\end{equation}
where $\eta$ and $\eta^{\prime}$ can independently take values +1, --1, or 
0.

	Since the convolution of distributions is not guaranteed to be a well
defined operation,$^{\ref{G-S62}}$ we have to examine all possible combinations
of $\eta$ and $\eta^{\prime}$. Actually, it is sufficient to consider the
cases $\eta=\pm 1$, $\eta^{\prime}=\pm 1$; other cases where $\eta=0$ or/and
$\eta^{\prime}=0$ reduce to the previous ones due to the linearity of
equations (A.2), (A.3) and the equality $D_0 = \ha(D_+ + D_-)$.

	Let us write down the expression (A.4) in the explicit form,
%
\begin{equation}
A(x) = \frac{1}{4\pi}\int\!\!d^4y\int\!\!d^4z\,D_\eta(y)
D_{\eta^{\prime}}(z)J(x-y-z).
\end{equation}
Representing $D_\eta$ in the form
%
\begin{equation}
D_\eta(x) = 2\Theta(\eta x^0)\delta(x^2) = \delta(x^0 - \eta|\B x|)/|\B x|,
\qquad \eta=\pm 1,
\end{equation}
where $\B x\equiv\{x^i, i=1,2,3\}$, yields for (A.5) the expression
%
\begin{equation}
A(x) = \frac{1}{4\pi}\int\!\!d^3y\int\!\!d^3z\,\frac{J(x^0-\eta|\B y|-
\eta^{\prime}|\B z|,\,\B{x-y-z})}{|\B y|\,|\B z|},
\end{equation}
or, in terms of new variables $\B u=\B{y+z}$, $\B v=\B{y-z}$,
%
\begin{eqnarray}
A(x) &=& \frac{1}{8\pi}\int\!\!d^3u\int\!\!d^3v\times \nonumber \\
&&\times\frac{J(x^0-\ha\eta|\B{u+v}|-
\ha\eta^{\prime}|\B{u-v}|,\,\B{x-u})}{|\B{u+v}|\,|\B{u-v}|}.
\end{eqnarray}

	Let us calculate the internal integral over $d^3v$ in r.h.s. of
(A.8). Expressing Cartesian coordinates $v_1$, $v_2$, $v_3$ of $\B v$
in terms of ellipsoidal coordinates,
%
\begin{eqnarray}
v_1 &=& |\B u|\sqrt{(\sigma^2-1)(1-\tau^2)}\cos\varphi, \nonumber \\
v_2 &=& |\B u|\sqrt{(\sigma^2-1)(1-\tau^2)}\sin\varphi, \nonumber \\
v_3 &=& |\B u|\sigma\tau,  \\
&& \sigma\ge1\ge\tau\ge-1,\qquad 0\le\varphi<2\pi, \nonumber
\end{eqnarray}
we obtain
%
\begin{eqnarray}
A(x) &=& \frac{1}{8\pi}\int\!\!d^3u\,|\B u|\int\limits_1^\infty\!\!d\sigma
\int\limits_{-1}^1\!\!d\tau\int\limits_0^{2\pi}\!\!d\varphi\times
\nonumber \\
&&\times J\left(x^0-\eta |\B u|\times\left\{\sigma,\ \ \eta^{\prime} = \eta\
\atop\tau,\ \ \eta^{\prime} = -\eta\right\},\,\B{x-u}\right).
\end{eqnarray}
If $\eta^{\prime}=-\eta$ this integral diverges due the factor
$\int_1^\infty\!\!d\sigma$. In the case $\eta^{\prime}=\eta$ it reads:
%
\begin{equation}
A(x) = \ha\int\!\!d^3u\,|\B u|\int\limits_1^\infty\!\!d\sigma
J(x^0-\eta |\B u|\sigma,\,\B{x-u}).
\end{equation}
Using the change of the variable $\sigma\to u^0=\eta|\B u|\sigma$ we obtain
the final expression for $A(x)$:
%
\begin{equation}
A(x) = \ha\int\!\!d^4u\,\Theta(\eta u^0)\Theta(u^2)
J(x-u).
\end{equation}
It follows from (A.12) that fundamental solutions to the equation (A.1)
are:
%
\begin{equation}
E_\eta(x) = \ha\Theta(\eta x^0)\Theta(x^2),\qquad
\eta=\pm 1.
\end{equation}
Since their supports are the interiors of future-- and past-oriented
light cones, these distributions are the retarded and the advanced
Green's functions of the equations (A.1). The time-symmetric Green's
function is constructed by linearity,
%
\begin{equation}
E_0(x) = \ha(E_+(x) + E_-(x)) = \qu\Theta(x^2).
\end{equation}
Eqs. (A.13), (A.14) are unified in eq. (21).

	We note that some complex fundamental solution to (A.1) is
obtained by means of another technique in Ref. \ref{G-S62},
%
\begin{equation}
E_c(x) = \frac{\pm i}{4\pi}\ln(x^2\pm i0).
\end{equation}
Its real part,
%
\begin{equation}
\Re E_c(x) = \qu(\Theta(x^2) - 1),
\end{equation}
coincides with the symmetric Green's function (A.15) up to a constant
(which is the solution of homogeneous equation). This solution can be
considered as the analogue of the Feynman propagator in QED,
%
\begin{equation}
D_c(x) = \frac{\pm i}{\pi(x^2\pm i0)},
\end{equation}
real part of which is the symmetric Green's function of d'Alembert equation.

\subsection*{B. Analysis of DPM for general model.}
\renewcommand{\theequation}{B.\arabic{equation}}
\setcounter{equation}{0}

Let us introduce the dimensionless positive quantities:
%
\begin{equation}
\xi = \frac{2\alpha}{Mr},\quad \mu = \frac{M}{2m},\quad \sigma =
\frac{S}{\alpha},\quad \nu = \frac{\alpha\beta}{m^2},
\end{equation}
and functions:
%
\begin{eqnarray}
\bar f_\pm(\xi,\mu,\sigma)&=&1 + 2\xi - \frac{\xi}{\xi-\nu/\mu^2}
\bar h_\pm(\xi,\mu,\sigma), \\
\bar h_\pm(\xi,\mu,\sigma)&=&\bar g_-(\xi,\mu,\sigma)
\mp\sqrt{\bar d(\xi,\mu,\sigma)}, \\
\bar d(\xi,\mu,\sigma)&=&\bar g_-^2(\xi,\mu,\sigma) +
\sigma^2\left(\xi - \frac{\nu}{\mu^2}\right)\xi^3, \\
\bar g_\pm(\xi,\mu,\sigma)&=&\frac{1}{2}\left[\frac{1}{\mu^2}(1+\xi) \pm
\left(\xi - \frac{\nu}{\mu^2}\right)\xi + \sigma^2\xi^2\right],
\end{eqnarray}
which are related to (70)--(73) as follows:
%
\begin{eqnarray}
\bar f_\pm(\xi,\mu,\sigma)&=&\frac{4}{M^2}f_\pm(r,M,S) \\
\bar h_\pm(\xi,\mu,\sigma)&=&\frac{4}{M^2}h_\pm(r,M,S) \\
\bar d(\xi,\mu,\sigma)&=&\frac{16}{M^4}d(r,M,S) \\
\bar g_-(\xi,\mu,\sigma)&=&\frac{4}{M^2}g(r,M,S)
\end{eqnarray}
Then DPM is determined by conditions:
%
\begin{equation}
\bar f_-(\xi,\mu,\sigma) \ge 0,
\end{equation}
%
%
\begin{equation}
\bar{\Delta} \equiv (1 + \xi)^2 - \bar f_-(\xi,\mu,\sigma) > 0.
\end{equation}
Although $\xi$ is positive by definition, it is useful to consider
the functions (B.2)--(B.5) of $\xi$ for $\xi\in\IR$.

	First of all we consider the condition (B.10). It follows from
(B.5) and (B.4) that
%
\begin{equation}
\bar g_-(\xi) > 0,\qquad
\xi\in[0,\nu/\mu^2],
\end{equation}
%
%
\begin{eqnarray}
\bar d(\xi) > \bar g_-^2(\xi),&&\qquad
\xi\in(-\infty,0)\cup(\nu/\mu^2,\infty), \\
\bar d(\xi) = \bar g_-^2(\xi),&&\qquad
\xi=0,\nu/\mu^2, \\
\bar d(\xi) < \bar g_-^2(\xi),&&\qquad
\xi\in(0,\nu/\mu^2).
\end{eqnarray}
Thus from (B.12)--(B.14) we obtain
%
\begin{equation}
\bar d(\xi) > 0,\qquad
\xi\in(-\infty,0]\cup[\nu/\mu^2,\infty).
\end{equation}
Using the equivalent form of $\bar d$,
%
\begin{equation}
\bar d(\xi,\mu,\sigma)=\bar g_+^2(\xi,\mu,\sigma) -
\frac{1}{\mu^2}(1+\xi)\left(\xi - \frac{\nu}{\mu^2}\right)\xi,
\end{equation}
we obviously have
%
\begin{eqnarray}
\bar d(\xi) > \bar g_+^2(\xi),&&\qquad
\xi\in(-\infty,-1)\cup(0,\nu/\mu^2), \\
\bar d(\xi) = \bar g_+^2(\xi),&&\qquad
\xi=-1,0,\nu/\mu^2, \\
\bar d(\xi) < \bar g_+^2(\xi),&&\qquad
\xi\in(-1,0)\cup(\nu/\mu^2,\infty),
\end{eqnarray}
and then, from (B.16) and (B.18),
%
\begin{equation}
\bar d(\xi) > 0,\qquad
\xi\in\IR.
\end{equation}
Hence the functions $\bar h_\pm(\xi)$, $\bar f_\pm(\xi)$ (as well as
$h_\pm(r)$, $f_\pm(r)$) are real.

	Taking into account (B.3), (B.12)--(B.15), and (B.21) one obtains
%
\begin{eqnarray}
\bar h_+(\xi) > 0,&&\qquad
\xi\in(0,\nu/\mu^2), \\
\bar h_+(\xi) = 0,&&\qquad
\xi=0,\nu/\mu^2, \\
\bar h_+(\xi) < 0,&&\qquad
\xi\in(-\infty,0)\cup(\nu/\mu^2,\infty)
\end{eqnarray}
Thus, using (B.2) and (B.22)--(B.24) we conclude that
%
\begin{equation}
\bar f_+(\xi) > 0,\qquad
\xi\in[-1/2,\infty).
\end{equation}
We note that the function $\bar f_+(\xi)$ is smooth at $\xi=\nu/\mu^2$.

	In order to clarify the behaviour of the function $\bar f_-(\xi)$
for $\xi>0$ let us consider the function $\bar f_+\bar f_-$. It can
be presented in the form:
%
\begin{equation}
\bar f_+(\xi)\bar f_-(\xi) = \frac{1+\xi}{\xi-\nu/\mu^2}\Pi(\xi),
\end{equation}
where
%
\begin{equation}
\Pi(\xi) \equiv (1+2\xi)\left[(1+\xi)\left(\xi-
\frac{\nu}{\mu^2}\right) - \frac{1}{\mu^2}\xi\right] -
\sigma^2(1+\xi)\xi^3
\end{equation}
is 4th-order polynomial (in terms of original quantities it is written
down in l.h.s. of (74)). It is evident that
%
\begin{equation}
\Pi(\xi) < 0 \qquad {\rm for}\ |\xi|\ {\rm large}.
\end{equation}
Moreover, it is easy to examine that
%
\begin{eqnarray}
\Pi(\xi) < 0,\ \ &&\qquad
\xi\in(-\infty,-1]\cup[0,\nu/\mu^2], \\
\Pi(-\ha) > 0.&& 
\end{eqnarray}
Thus $\Pi(\xi)$ has two negative roots, $\xi_1$ and $\xi_2$, $-1<\xi_1<-1/2<
\xi_2<0$, which exist at arbitrary (positive) values of $\mu$, $\sigma$, and
$\nu$. The number of positive roots, should they exist, is not more than two.
Let us note that one can choose the sufficiently large value of $\sigma$ such
that $\Pi(\xi)>0$, $\xi>0$. On the other hand, at $S=0$ there exists
$\xi_+>0$ such that $\Pi(\xi)>0$, $\xi>\xi_+$. Thus, given $\mu$ and $\nu$,
two other roots, $\xi_3$ and $\xi_4$, are positive for sufficiently small
values of $\sigma$, and $\nu/\mu^2<\xi_3<\xi_4$. In this case we have:
%
\begin{eqnarray}
\Pi(\xi) > 0,&&\qquad
\xi\in(\xi_1,\xi_2)\cup(\xi_3,\xi_4), \\
\Pi(\xi) < 0,&&\qquad
\xi\in(-\infty,\xi_1)\cup(\xi_2,\xi_3)\cup(\xi_4,\infty).
\end{eqnarray}

	Hereafter we restrict all functions on $\xi>0$. Using (B.26),
(B.31), (B.32), and (B.25) one concludes that
%
\begin{eqnarray}
\bar f_-(\xi) \ge 0,&&\qquad
\xi\in(0,\nu/\mu^2)\cup[\xi_3,\xi_4], \\
\bar f_-(\xi) < 0,&&\qquad
\xi\in(\nu/\mu^2,\xi_3)\cup(\xi_4,\infty),
\end{eqnarray}
and this function has a pole at $\xi=\nu/\mu^2$.

	Now let us consider the condition (B.11) which can be presented in
the following form:
%
\begin{equation}
\bar{\Delta} = \frac{\xi}{\xi-\nu/\mu^2}
\left[\bar g_+(\xi,\mu,\sigma)
+\sqrt{\bar d(\xi,\mu,\sigma)}\right] > 0.
\end{equation}
Using (B.18), (B.20) and the evident inequality
%
\begin{equation}
\bar g_+(\xi) > 0,\qquad \xi\in(\nu/\mu^2,\infty),
\end{equation}
one concludes that
%
\begin{eqnarray}
\bar{\Delta} > 0,&&\qquad
\xi\in(\nu/\mu^2,\infty), \\
\bar{\Delta} < 0,&&\qquad
\xi\in(0,\nu/\mu^2).
\end{eqnarray}

	Finally, taking into account (B.33), (B.34), (B.37), and (B.38),
we find DPM:
$\xi_3\le\xi\le\xi_4$, i.e., $r_1\le r\le r_2$, where $r_1=2\alpha/(M\xi_4)$,
$r_2=2\alpha/(M\xi_3)$. It disappear if $\xi_3\to\xi_4$. The degenerated case
$\xi_3=\xi_4\equiv\xi_c$, where $\xi_c$ satisfies the set of equations:
%
\begin{equation}
\Pi(\xi_c)= 0,\qquad \Pi^\prime(\xi_c)=0,
\end{equation}
corresponds to circular orbit motion.  The set (B.39) in this case can be
considered as a relation between $M$ and $S$ which is presented by (75),
where $\lambda=1/\xi_c$.


\section*{References}

\normalsize
\begin{enumerate}
\item\label{Sim94}
Yu. A. Simonov, {\em Nuovo Cimento}, {\bf A107}, 2629 (1994). 
\item\label{S-F88}
A. Slavnov and L. Faddeev, {\em Introduction to quantum theory of gauge 
fields}, (Nauka, Main Editorial Board for Physical and Mathematical 
Literature, Moskow, 1988) (in Russian).
\item\label{L-S89}
W. Lucha and F. F. Sch\"oberl, {\em Die Starke Wechselwirkung. Eine 
Einf\"uhrung in nichtrelativistische Potentialmodelle} (Bibliographishches
Institut \& F. A. Brockhaus, Mannheim, 1989).
\item\label{FKR71}
R. P. Feynman, N. Kislinger and F. Ravndel, {\em Phys. Rev.} {\bf D3}, 2706
(1971).
\item\label{Ono81}
S. Ono, {\em Phys. Rev.} {\bf D26}, 2510 (1981).
\item\label{GCO89}
C. Goebel, D. LaCourse and M. G. Olsson, {\em Phys. Rev.} {\bf D41}, 2917 
(1989).
\item\label{HLS94}
I. I. Haysak, V. I. Lengyel and A. O. Shpenik, in  {\em Hadrons-94. Proc. of
Workshop on Soft Physics (Strong Interaction at Large Distance)}, Uzhgorod,
1994, eds. G. Bugrij, L. Jenkovszky and E. Martynov 
(Bogoliubov Institute for Theoretical Physics, Kiev, 1994), pp. 267--271.;
S. Chalupka, V. I. Lengyel, M. Salak and A. O. Shpenik, {\em ibid.}, 
pp. 272--279.
\item\label{Bar82}
J. Llosa (ed.), {\em Relativistic  Action  at a Distance: Classical and 
Quantum Aspects. Proc. of Workshop}, Barcelona, 1981, Lectures Notes in
Physics, No~162 (Springer, Berlin, 1982).
\item\label{GKT87}
R. P. Gaida, Yu. B. Kluchkovsky and V. I. Tretyak, in Ref. \ref{Fir87}, 
pp.~210--241.
\item\label{L-S78}
H. Leutwyler and J. Stern, {\em Ann. Phys. (N.Y.)} {\bf 112}, 94 (1978).
\item\label{Pol89}
W. N. Polyzou, {\em Ann. Phys. (N.Y.)} {\bf 193}, 367 (1989).
\item\label{Fir87}
G. Longhi and L. Lusanna (eds.), {\em Constraint's Theory and Relativistic
Dynamics. Proc. Workshop}, Firenze, 1986 (World Scientific, Singapore, 1987).
\item\label{H-N74}
F. Hoyle and J. V. Narlikar, {\em Action-at-a-Distance in Physics and 
Cosmology} (W. H. Freeman \& Co., San Francisco, 1974).
\item\label{Tak79}
T. Takabayasi, {\em Supp. Progr. Theor. Phys.} {\bf 67}, 1 (1979).
\item\label{Roh79}
F. Rohrlich, {\em Physica} {\bf A96}, 290 (1979);
M. King and F. Rohrlich, {\em Phys. Rev. Lett.} {\bf 44}, 621 (1980);
T. Biswas and F. Rohrlich, {\em Nuovo Cimento} {\bf A88}, 125 (1984);
T. Biswas, {\em ibid}, 145;
H. W. Crater and P. Van Alstine, {\em Phys. Rev. Lett.} {\bf 53}, 1527 (1984);
in Ref. \ref{Fir87}, pp.~171--195;
H. Sazdjian, {\em Phys. Rev.} {\bf D33} 3425 (1985);
D. D. Brayshaw, {\em Phys. Rev.} {\bf D36}, 1465 (1986);
N. A. Aboud and J. R. Hiller, {\em Phys. Rev.} {\bf D41}, 937 (1990);
S. Ishida and M. Oda, {\em Nuovo Cimento} {\bf A107}, 2519 (1994).
\item\label{Hav71}
P. Havas, "Galilei- and Lorentz-invariant particle  systems  and
their conservation laws", in {\em Problems  in  the  Foundations  of
Physics}, ed. M. Bunge (Springer, Berlin, 1971), pp.~31--48;
P. Ramond, {\em Phys. Rev.} {\bf D7}, 449 (1973);
V. I. Tretyak, Second PhD degree thesis, Lviv State University, 1996
(in Ukrainian); Institute for Condensed Matter Physics, Lviv, 
Preprint ICMP--98--03U, 1998 (in Ukrainian; unpublished\footnote{On the 
WEB: http://www.icmp.lviv.ua/}).
\item\label{A-A84}
A. I. Alekseev and B. A. Arbuzov, {\em Theor. Math. Phys.} {\bf 59}, 372 
(1984).
\item\label{AAB82}
A. I. Alekseev, B. A. Arbuzov and V. A. Baykov, {\em Theor. Math. Phys.}
{\bf 52}, 739 (1982).
\item\label{Ale88}
A. I. Alekseev, Institute for High Energy Physics, Sepukhov, Preprint IHEP 
88--97, 1988 (unpublished);
A. I. Alekseev and B. A. Arbuzov, 
{\em Phys. Lett.} {\bf B242}, 103 (1990).
\item\label{Riv84}
A. Rivacoba, {\em Nuovo Cimento} {\bf B84}, 35 (1984).
\item\label{Wei86}
J. Weiss, {\em J.~Math. Phys.} {\bf 27}, 1015 (1986).
\item\label{D-T93}
A. A. Duviryak and V. I. Tretyak, {\em Condensed Matter Physics} {\bf 1},
92 (1993) (in Ukrainian);
A. Duviryak, {\em Acta Phys. Polon.} {\bf B28}, 1087 ( 1997).
\item\label{Won70}
S. K. Wong, {\em Nuovo Cimento} {\bf A65}, 689 (1970).
\item\label{dGS72}
S. R. de Groot and L. G. Suttorp, {\em Foundations of electrodynamics}
(North--Holland Publishing Company,  Amsterdam, 1972).
\item\label{Kat69}
A. Katz, {\em J.~Math. Phys.} {\bf 10}, 1929 (1969).
\item\label{Whe49}
J. A. Wheeler and R. P. Feynman, {\em Rev. Mod. Phys.} {\bf 21}, 425 (1949).
\item\label{Sta70}
A. Staruszkiewicz, {\em Ann. der Physik} {\bf 25}, 362 (1970);
R. A. Rudd and R. N. Hill, {\em J. Math. Phys.} {\bf 11}, 2704 (1970).
\item\label{Kun74}
H. P. K\"unzle, {\em Int. J. Theor. Phys.} {\bf 11}, 395 (1974);
D. E. Fahnline, {\em J. Math. Phys.} {\bf 22} 1640 (1981).
\item\label{B-T53}
B. Bakamjian and L. H. Thomas, {\em Phys. Rev.} {\bf 92}, 1300 (1953).
\item\label{D-K92}
A. A. Duviryak and Yu. B. Kluchkovsky, {\em Ukr. Fiz. Zh.} {\bf 37}, 313 
(1992) (in Ukrainian);
{\em J. Soviet. Math.} {\bf 66}, 2648 (1993).
\item\label{Shp98}
V. Shpytko, Institute for Condensed Matter Physics, Lviv, 
Preprint ICMP--98--12E, 1998 (unpublished$^c$).
\item\label{G-S62}
I. M. Guelfand and G. E. Shilov, {\em Generalized Functions} (Academic, 1964),
Vol. 1.
\end{enumerate}
\end{document}